\documentclass[aps,showpacs,superscriptaddress]{revtex4}
\usepackage{amsmath}
\usepackage{graphicx}

\begin{document}

\title{Entanglement production and decoherence-free subspace of two
single-mode cavities embedded in a common environment}
\author{Jun-Hong An}
\affiliation{Department of Modern Physics, Lanzhou University, 730000 Lanzhou, China}
\author{Shun-Jin Wang\footnote{
E-mail address of the corresponding author: sjwang@home.swjtu.edu.cn} }
\affiliation{Department of Modern Physics, Lanzhou University, 730000 Lanzhou, China}
\affiliation{Department of Physics, Sichuan University, 610064 Chengdu, China}
\author{Hong-Gang Luo}
\affiliation{Department of Modern Physics, Lanzhou University, 730000 Lanzhou, China}
\affiliation{Institute of Theoretical Physics, Chinese Academy of Sciences, 100080
Beijing, China}

\begin{abstract}
A system consisting of two identical single-mode cavities coupled to a
common environment is investigated within the framework of algebraic
dynamics. Based on the left and right representations of the Heisenberg-Weyl
algebra, the algebraic structure of the master equation is explored and
exact analytical solutions of this system are obtained. It is shown that for
such a system, the environment can produce entanglement in contrast to its
commonly believed role of destroying entanglement. In addition, the
collective zero-mode eigen solutions of the system are found to be free of
decoherence against the dissipation of the environment. These
decoherence-free states may be useful in quantum information and quantum
computation.
\end{abstract}

\maketitle

\section{Introduction}

Quantum entanglement plays a basic role in quantum communication and quantum
computation \cite{Ekert}. The creation of entanglement between qubits is of
fundamental importance for further quantum computation processing. The
entanglement can be created by a direct interactions between qubits \cite%
{Peter} or an indirect interactions via a third party \cite{Bose}.
However, both of the above processes are confined in closed
system, i.e., the influences of environment are neglected. Recent
investigations showed that environment can be helpful to the
entanglement creation in an open system \cite{Braun, Benatti},
which provide a perspective to use the environment to implement
decoherence-free quantum information processing \cite{Beige,
Bouren}. In order to treat the influences of the environment on an
open quantum system, the Born-Markovian master equation approach
has been widely used. The common feature of the quantum master
equations is the existence of the sandwich terms of the Liouville
operator where the reduced density matrix of the system is in
between some quantum excitation and de-excitation operators. These
terms result from the elimination of an enormous irrelevant
degrees of freedom of the environment. Except for some simple
cases, for example, a single-mode of the cavity field coupled to
the vacuum or stationary regime properties \cite{Scu}, it is very
difficult to solve directly the master equation. Instead, it is
usual to convert the master equations into some c-number equations
in the coherent state representation--the Fokker-Planck equation
\cite{Gar,Wall}.

In the previous works \cite{Wang01,An,Zhao}, we have proposed and developed
an algebraic method to treat the sandwich terms in the Liouville operator
for quantum statistical systems. This method is just a generalization of the
algebraic dynamical method \cite{Wang93} from quantum mechanical systems to
quantum statistical systems. According to the characteristic of the sandwich
terms in the Liouville operator, the left and right representations of the
relevant algebra \cite{Wang89} have been introduced and the corresponding
composite algebra has been constructed. As a result, the master equation has
been converted into a Schr\"{o}dinger-like equation and the problems can be
solved exactly.

In this paper, we shall use this method to solve the problem of two
identical cavities coupled to a common environment. The system consisting of
two coupled cavities and a similar system of two coupled harmonic
oscillators are important in quantum optics and quantum information theory.
From these systems, many properties, such as, information transfer of
quantum states \cite{Dodonov}, and the quantum statistical properties of the
two-coupled modes of electromagnetic fields \cite{Kalmykov}, have been
investigated. In Ref. \cite{Dodonov2002}, the entanglement of the two
coupled harmonic oscillator system has been quantitatively studied. However,
all these studies neglect the effect of environment on the modes of cavities
and the system is thus closed. Although Ref. \cite{ZhaoY} studied the
dissipative two oscillators system, it was based on the quantum
characteristic function approach and the direct analytical solutions are not
obtained. In this work, we shall consider a specific coupling of the two
cavity fields, induced through the individual interactions of the two
cavities with the environment and our system is thus different from Ref.
\cite{ZhaoY}. By introducing a collective mode consisting of the two cavity
modes, the quantum master equation is obtained. The two independent cavities
are thus interacting with each other indirectly through the environment.
Using the algebraic dynamical method, the full algebraic structure of the
master equation is explored and its exact analytical solution is obtained.
From the zero eigenvalue solutions, the decoherence-free sub-space is
obtained. Finally, the exact entanglement dynamics for the single- and
two-photon processes is obtained from the time-dependent solutions. We find
that the system exhibits an interesting feature of environment-assisted
entanglement generation. Our present study shows that the environment is not
just a negative source of decoherence, it also can be a positive source of
entanglement.

The paper is organized as follows. In Sec. II, the model Hamiltonian of the
system is presented and the master equation for the reduced density matrix
of the system is obtained. In Sec. III, based on the left and right
representations of the relevant algebra, we introduce a composite algebra,
in terms of which the dynamical $su(1,1)$ plus $u(2)$ algebraic structures
of the Liouville operator( rate operator ) of the master equation are found.
Sec. IV is devoted to the analytical solutions of the master equation. The
applications of its time-dependent solutions to the entanglement dynamics
for the single- and two-photon processes are studied in Sec. V. Finally, a
brief summary is given in Sec. VI.

\section{\protect\bigskip Two identical cavities in a common environment}

Consider two identical cavities (or two harmonic oscillators) interacting
with a common environment. With the dipole interaction and in the rotating
wave approximation, the system can be described by the Hamiltonian
\begin{equation*}
H=\omega \sum_{i}a_{i}^{\dagger }a_{i}+\sum_{k}\omega _{k}b_{k}^{\dagger
}b_{k}+\sum_{i,k}(g_{ik}a_{i}^{\dagger }b_{k}+H.c.)
\end{equation*}%
where $a_{i}(i=1,2)$ are boson operators of the two cavities with the same
frequency $\omega $ and $b_{k}$ are those for the environment fields,
respectively. $g_{ik}$ is the coupling constant. It is reasonable to assume
that the coupling of the two cavities to the environment is same, i.e., $%
g_{ik}=g_{k}$. This is a generation of Dicke limit of quantum optics \cite%
{Hep}. Using the standard technique of quantum optics, one can obtain the
master equation for the reduced density of the two-cavity system under the
standard Born-Markovian approximation \cite{Wall,Scu}
\begin{eqnarray}
\frac{d\rho (t)}{dt} &=&\frac{\varsigma }{2}\{[2a_{1}\rho (t)a_{1}^{\dagger
}-a_{1}^{\dagger }a_{1}\rho (t)-\rho (t)a_{1}^{\dagger }a_{1}]+[2a_{2}\rho
(t)a_{2}^{\dagger }-a_{2}^{\dagger }a_{2}\rho (t)-\rho (t)a_{2}^{\dagger
}a_{2}]\}  \notag \\
&&+\frac{\varsigma }{2}\{[2a_{1}\rho (t)a_{2}^{\dagger }-a_{1}^{\dagger
}a_{2}\rho (t)-\rho (t)a_{1}^{\dagger }a_{2}]+[2a_{2}\rho (t)a_{1}^{\dagger
}-a_{2}^{\dagger }a_{1}\rho (t)-\rho (t)a_{2}^{\dagger }a_{1}]\},
\label{masa}
\end{eqnarray}%
where $\varsigma (>0)$ is the decay constant of the collective
mode. It is noted that on the right-hand side of Eq. (\ref{masa})
the first two terms denote the individual dissipation of the two
cavities due to the environment, while the last two terms describe
the coupling (photon
exchange) between the two cavities mediated by the environment. Eq. (\ref%
{masa}) is complicated and can be simplified by introducing a collective
(quasi-photon) mode operator $A=\frac{1}{\sqrt{2}}(a_{1}+a_{2})$, as did in
Ref. \cite{Zan}. In doing so, the master equation of the reduced density
becomes
\begin{equation}
\frac{d\rho (t)}{dt}=\frac{\varsigma }{2}[2A\rho (t)A^{\dagger }-A^{\dagger
}A\rho (t)-\rho (t)A^{\dagger }A].  \label{mas}
\end{equation}%
In the following we shall discuss the algebraic structure of this system.

\section{\protect\bigskip Algebraic structure of the master equation}

Based on the left and right representations of Heisenberg-Weyl algebra \cite%
{Wang01}, the master equation (\ref{mas}) can be converted into a Schr\"{o}%
dinger-like equation and the algebraic structure can thus be established.
Noticing the fact that the collective operators $A$ and $A^{\dagger }$ obey
the same commutation relations as the photon operators $a$ and $a^{\dagger }$
do, $\{A,A^{\dagger },A^{\dagger }A,1\}$ thus constitutes a $hw(4)$\
algebra. Define the left and right algebras of $hw(4)$%
\begin{eqnarray}
hw(4)_{r} &=&\{A^{r},A^{r\dagger },\tilde{N}^{r}=A^{r\dagger }A^{r},1\},
\notag \\
hw(4)_{l} &=&\{A^{l},A^{l\dagger },\tilde{N}^{l}=A^{l\dagger }A^{l},1\},
\label{alg}
\end{eqnarray}
where $hw(4)_{r}$ acts to the right on the ket photon number state $%
|n\rangle $ and $hw(4)_{l}$ acts to the left on the bra photon number state $%
\langle n|$. They have the following commutators
\begin{eqnarray}
\lbrack A^{r},A^{r\dagger }] &=&1,[A^{r},\tilde{N}^{r}]=A^{r},[A^{r\dagger
}, \tilde{N}^{r}]=-A^{r\dagger };  \notag \\
\lbrack A^{l},A^{l\dagger }] &=&-1,[A^{l},\tilde{N}^{l}]=-A^{l},[A^{l\dagger
},\tilde{N}^{l}]=A^{l\dagger }.  \label{commut}
\end{eqnarray}
It is noted that $hw(4)_{l}$ ($hw(4)_{r}$) is isomorphic (anti-isomorphic)
to $hw(4)$. Since $hw(4)_{l}$ and $hw(4)_{l}$ act on different spaces-bra
and ket spaces, the operators commute with each other, i.e., $%
[hw(4)_{l},hw(4)_{r}]=0$. The reduced density matrix is a vector of the von
Nuemann super space, which has the $hw(4)_{l}\otimes hw(4)_{r}$ algebraic
structure containing $sp(4)$ as a relevant sub-algebra, as shown below.

From the above basic algebras we can constitute a composite algebra $C$ in
the adjoint representation,
\begin{equation*}
C=\{K_{-}=A^{r}A^{l\dagger },K_{+}=A^{r\dagger }A^{l},K_{0}=\frac{1}{2}(%
\tilde{N}^{r}+\tilde{N}^{l})\}
\end{equation*}%
We see that $C$ is an $su(1,1)$ algebra satisfying the following
commutation rules
\begin{equation*}
\lbrack K_{0},K_{\pm }]=\pm K_{\pm },[K_{-},K_{+}]=2K_{0},
\end{equation*}%
which is an sub-algebra of the $sp(4)$ and can be derived from Eqs. (\ref%
{commut}). The operator $K_{-}$ acts on the photon number bases of the von
Neumann space as follows
\begin{eqnarray}
K_{-}|n_{1}n_{2}\rangle \langle m_{1}m_{2}| &=&\frac{1}{2}(\sqrt{n_{1}}%
|n_{1}-1,n_{2}\rangle +\sqrt{n_{2}}|n_{1},n_{2}-1\rangle )  \notag \\
&&(\langle m_{1}-1,m_{2}|\sqrt{m_{1}}+\langle m_{1},m_{2}-1|\sqrt{m_{2}}).
\label{ac1}
\end{eqnarray}%
Similarly, for the super-vector bases of the quasi-photon (collective
photon) number states $|n\rangle _{qq}\langle m|$ marked by the subscript $q$%
, the actions of the de-excitation operator $K_{-}$ and the number operator $%
K_{0}$ read
\begin{eqnarray}
K_{-}|n\rangle _{qq}\langle m| &=&\sqrt{nm}|n-1\rangle _{qq}\langle m-1|,
\notag \\
K_{0}|n\rangle _{qq}\langle m| &=&\frac{n+m+1}{2}|n\rangle
_{qq}\langle m|. \label{action}
\end{eqnarray}

With this composite algebra at hand, it is straightforward to convert the
master equation (\ref{mas}) into Schr\"{o}dinger-like equation
\begin{eqnarray}
\frac{d\rho (t)}{dt} &=&\Gamma \rho (t),  \notag \\
\Gamma &=&\varsigma K_{-}-\varsigma K_{0}+\frac{\varsigma }{2}.  \label{mast}
\end{eqnarray}
Since the rate operator $\Gamma $ is a linear function of the
$su(1,1)$ generators, we conclude that the master equation
possesses an $su(1,1)$ dynamical symmetry in the quasi-photon
number representation. Thus the system is integrable and can be
solved analytically according to the algebraic dynamics
\cite{Wang93}.

It is remarkable to note that $K_{0}$ is diagonalized in the quasi-photon
representation, but it is not so in the real photon number representation of
the von Neumann space of the density matrix. As will be seen soon, it has a
different algebraic structure in the real photon number representation. In
terms of the two real photon operators, $K_{0}$ can be written as
\begin{equation*}
K_{0}=\frac{1}{4}(N_{1}+N_{2}+S_{+}+S_{-}),
\end{equation*}
where $N_{i}=a_{i}^{r\dagger }a_{i}^{r}+a_{i}^{l\dagger
}a_{i}^{l}=n_{i}^{r}+n_{i}^{l}(i=1,2)$, $S_{+}=a_{1}^{r\dagger
}a_{2}^{r}+a_{1}^{l}a_{2}^{l\dagger }$, and $S_{-}=a_{1}^{r}a_{2}^{r%
\dagger}+a_{1}^{l\dagger }a_{2}^{l}$. It is not difficult to find that the
operators $(N_{1},N_{2},S_{+},S_{-})$ form a $u(2)(=u(1)+su(2))$ algebra
\begin{equation*}
u(2)=\{S_{0}=(N_{1}-N_{2})/2,S_{+},S_{-};N=(N_{1}+N_{2})/2\},
\end{equation*}
which obey the following commutation rules
\begin{eqnarray*}
&&[N,S_{\pm }]=[N,S_{0}]=0, \\
&&[S_{0},S_{\pm }]=\pm S_{\pm },[S_{-},S_{+}]=-2S_{0}.
\end{eqnarray*}
These operators act on the von Neumann space in the real photon number
representation as follows:
\begin{eqnarray}
S_{+}|n_{1}n_{2}\rangle \langle m_{1}m_{2}| &=&\sqrt{(n_{1}+1)n_{2}}
|n_{1}+1,n_{2}-1\rangle \langle m_{1}m_{2}|+\sqrt{(m_{1}+1)m_{2}}
|n_{1}n_{2}\rangle \langle m_{1}+1,m_{2}-1|,  \notag \\
S_{-}|n_{1}n_{2}\rangle \langle m_{1}m_{2}| &=&\sqrt{n_{1}(n_{2}+1)}
|n_{1}-1,n_{2}+1\rangle \langle m_{1}m_{2}|+\sqrt{m_{1}(m_{2}+1)}
|n_{1}n_{2}\rangle \langle m_{1}-1,m_{2}+1|,  \notag \\
N_{i}|n_{1}n_{2}\rangle \langle m_{1}m_{2}|
&=&(n_{i}+m_{i}+1)|n_{1}n_{2}\rangle \langle m_{1}m_{2}|,(i=1,2)  \notag \\
S_{0}|n_{1}n_{2}\rangle \langle m_{1}m_{2}|
&=&(n_{1}+m_{1}-n_{2}-m_{2})/2|n_{1}n_{2}\rangle \langle m_{1}m_{2}|,  \notag
\\
N|n_{1}n_{2}\rangle \langle m_{1}m_{2}|
&=&(n_{1}+m_{1}+n_{2}+m_{2}+2)/2|n_{1}n_{2}\rangle \langle m_{1}m_{2}|.
\label{ac2}
\end{eqnarray}

Up to now we have explored the complete dynamical symmetry of the system and
the results can be summarized as follows: the largest dynamical algebra of
the two-cavity system is the $hw(4)_{l}\otimes hw(4)_{r}$ $\supset sp(4)$
algebra since it has two kinds of photons; due to the special structure of
the rate operator $\Gamma ,$ the system in fact has the$\ su(1,1)$ and $u(2)$
sub-algebraic dynamical symmetries of the largest dynamical symmetry algebra
$sp(4)$. For the quasi-photon, the dynamical symmetry is of the $su(1,1)$
sub-algebra and for the two kinds of real photons it is $u(2)$ algebra.

\section{Solutions of the master equation}

\subsection{Eigen solution and decoherence-free states}

To better understand the property of steady solution of the master equation,
we first investigate the eigensolutions of the master equation. The eigen
equation reads
\begin{equation}
\Gamma \rho =\gamma \rho .  \label{eig}
\end{equation}
To solve Eq. (\ref{eig}), we introduce two similarity transformations
corresponding to the above two sub-algebraic structures. The first one is
\begin{equation*}
\rho =U_{1}\rho _{1},U_{1}=e^{-K_{-}}.
\end{equation*}
The operator $\Gamma $ is thus diagonalized in the quasi-photon number
representation as follows
\begin{equation}
\bar{\Gamma}_{1}=U_{1}^{-1}\Gamma U_{1}=-\varsigma K_{0}+\frac{\varsigma }{2}%
.  \label{bar}
\end{equation}
From Eq. (\ref{bar}) the zero-mode eigensolution of $\Gamma $ in the
quasi-photon number representation can be obtained
\begin{equation}
\rho _{0}=|0\rangle _{qq}\langle 0|,  \label{steady}
\end{equation}
which has the same form as the stationary solution of the master equation
for the single-mode damped harmonic oscillator \cite{Scu}.

One notes that in the quasi-photon number representation, the collective
quasi-photon mode is used, in which two cavities are coupled. In
experiments, it is convenient to use the real photon number representation.
Moreover, in this representation, the interaction of the two cavities can
also be studied.

To diagonalize the rate operator $\bar{\Gamma}_{1}$(or $K_{0}$) in the real
photon number representation, the second similarity transformation is
needed, namely,
\begin{equation}
\bar\Gamma_2 = U_2^{-1} \bar\Gamma_1 U_2,
\end{equation}
where $U_2$ is defined as
\begin{equation*}
\rho _{1}=U_{2}\rho _{2},U_{2}=e^{\beta _{+}S_{+}}e^{\beta _{-}S_{-}}.
\end{equation*}
Under the conditions
\begin{eqnarray}
1-\beta _{+}^{2} &=&0,  \notag \\
2\beta _{+}\beta _{-}+1 &=&0,  \label{cond2}
\end{eqnarray}
i.e., $\beta _{+}=\pm 1$ and $\beta _{-}=\mp \frac{1}{2}$, the rate operator
$\bar{\Gamma}_{2}$ is diagonalized in terms of both of the two Cartan
operators $N$ and $S_{0}$ of the $u(2)$ algebra and the real photon number
representation$(N_{1},N_{2})$,
\begin{eqnarray*}
\bar{\Gamma}_{2} &=&\frac{-\varsigma }{2}(N-\beta _{+}S_{0})+\frac{\varsigma
}{2}, \\
&=&\frac{-\varsigma }{4}(f_{2}N_{1}+f_{3}N_{2})+\frac{\varsigma }{2}, \\
f_{2} &=&1-\beta _{+}, \\
f_{3} &=&1+\beta _{+}.
\end{eqnarray*}
From the expression of $\bar{\Gamma}_{2}$ we can see that the eigen
solutions of $\bar{\Gamma}_{2}$ are highly degenerate with respect
to different photon number distributions among the bra and ket
states. So besides the two Cartan operators we need two additional
quantum numbers to distinguish the degenerate states. Then the
complete set of commutation quantum operators is
$\{N,S_{0},n_{1}^{r},n_{2}^{r}\}$. For convenience, we
choose the equivalent set of the complete commutation quantum operators as $%
\{n_{1}^{r},$ $n_{1}^{l},n_{2}^{r},n_{2}^{l}\}$.

After making two inverse transformations of $\bar{\Gamma}_{2}$ and $\bar{
\Gamma}_{1}$, we obtain the eigen solution of Eq. (\ref{eig}) on the real
photon number bases of the von Neumann space as follows,
\begin{eqnarray}
\gamma (n_{1}n_{2;}m_{1}m_{2}) &=&\frac{-\varsigma }{4}
[f_{2}(n_{1}+m_{1}+1)+f_{3}(n_{2}+m_{2}+1)]+\frac{\varsigma }{2},  \notag \\
\rho (n_{1}n_{2;}m_{1}m_{2}) &=&e^{-K_{-}}e^{\beta _{+}S_{+}}e^{\beta
_{-}S_{-}}|n_{1}n_{2}\rangle \langle m_{1}m_{2}|.\ \   \label{eignest}
\end{eqnarray}
Substituting the two sets of solutions of Eqs. (\ref{cond2}) into the
expression of $\gamma $, we see that both of them contain the quasi-photon
zero-mode solution. For the second similarity transformation, the two sets
of solutions of Eqs. (\ref{cond2}) are physically equivalent and degenerate.
This can be seen from the expressions of $\gamma $, $f_{2}$, and $f_{3}$ in
terms of the transformation coefficient $\beta _{+}$ which produces the
solutions labeled by $n_{1}$ and $m_{1}$ (or $n_{2}$ and $m_{2}$), since $%
f_{2}$ and $f_{3}$ cannot be non-zero simultaneously under the similarity
transformation. The degeneracy of the zero quasi-photon mode comes from the
different real photon number distributions among the two cavities. From Eqs.
(\ref{eignest}) we can express the zero quasi-photon mode solution of the
system in terms of the real photon number bases as follows
\begin{equation}
\rho _{0}=c_{0}e^{-K_{-}}e^{S_{+}}e^{-\frac{1}{2}S_{-}}|n0\rangle \langle
m0|=c_{0}^{\prime }e^{-K_{-}}e^{-S_{+}}e^{\frac{1}{2}S_{-}}|0n\rangle
\langle 0m|  \label{stso}
\end{equation}
where $c_{0}$ and $c_{0}^{\prime }$ are normalized constants.
After some straightforward calculations, it is found that the
action of $e^{-K_{-}}$ have no effect on the states
$e^{S_{+}}e^{-\frac{1}{2}S_{-}}|n0\rangle \langle m0|$ (or
$e^{-S_{+}}e^{\frac{1}{2}S_{-}}|0n\rangle \langle 0m|$). Since
$\Gamma\rho_{0}=0$, $\rho_{0}$ are an invariant subspace under the
time evolution of the master equation (\ref{mast}). It is found
that the quasi-photon zero-mode subspace (\ref{stso}) contains
many entangled real photon states in the two cavities, which are
thus decohence-free with respect to the action of the dissipative
operator $\Gamma$ in the master equation (\ref{mast}). For
example, for the zero-photon process, the zero-mode state is $\rho
_{0}^{(0)}=|00\rangle \langle 00|$; for the single-photon process,
the zero-mode state is $\rho _{0}^{(1)}=|\phi \rangle \langle \phi
|$, where $|\phi \rangle =\frac{1}{\sqrt{2}}(|10\rangle
-|01\rangle )$; for the two-photon process, the zero-mode state is
$\rho _{0}^{(2)}=|\varphi \rangle \langle \varphi |$, where
$|\varphi \rangle = \frac{1}{2}(|02\rangle +|20\rangle
-\sqrt{2}|11\rangle )$, etc. All the above entangled real photon
states are the members of the quasi-photon zero-mode subspace and
decoherence-free from the dissipative action of $\Gamma$ in the
master equation(\ref{mast}).

From the above analysis, we see that the quasi-photon zero-mode solution (%
\ref{stso}) forms a highly degenerate and dissipation-free collective
subspace in terms of different real photon number states which are
orthogonal to each other, and many of them are entangled and thus
decoherence-free. It is noted that the equilibrium steady solutions of the
system depending on initial states are not unique, as shown later. All of
them are within the quasi-photon zero-mode subspace and consisting of all
possible mixture of these real photon states with collective zero-mode.

\subsection{Time-dependent solutions of the master equation}

Next we shall investigate the dynamical properties of the system based on
the real photon number representation. First we solve the master equation (%
\ref{mast}) and get its time-dependent solutions. Then we analyse the time
evolution behavior of the solutions and prove that for some initial product
states the entanglement can be produced by the environment induced dynamics
and for some initial state the system is decoherence-free.

By introducing a time-dependent gauge transformation
\begin{equation*}
\rho =U_{1}(t) \tilde{\rho}, \;\; U_{1}(t) =e^{\alpha _{-}(t)K_{-}},
\end{equation*}
the master equation can be rewritten as a diagonal form in the quasi-photon
number representation
\begin{eqnarray*}
\frac{d\tilde{\rho}(t)}{dt} &=&\tilde{\Gamma}^{\prime}\tilde{\rho}(t), \\
\tilde{\Gamma}^{\prime } &=&U_{1}^{-1}(t) \Gamma U_{1}(t) -U_{1}^{-1}(t)
\frac{dU_{1}(t) }{dt}=-\varsigma K_{0}+\frac{\varsigma }{2},
\end{eqnarray*}
if the following gauge transformation conditions are satisfied
\begin{equation}
\frac{d\alpha _{-}(t)}{dt}=\varsigma \lbrack 1+\alpha _{-}(t)].
\label{condit2}
\end{equation}
The time-dependent solution is now
\begin{eqnarray}
\rho (t) &=&U_{1}(t) \tilde{\rho}(t)=U_{1}(t) e^{\int_{0}^{t}\tilde{\Gamma}
(\tau )d\tau }\rho (0)  \notag \\
&=&\sum_{m,n}C_{m,n}e^{\alpha _{-}(t)K_{-}}e^{\frac{-\varsigma (n+m)t}{2}
}|n\rangle _{qq}\langle m|,  \label{solution}
\end{eqnarray}
where we have used the initial super state as follows $\rho (0)=$ $\tilde{%
\rho}(0)=\sum_{m,n}C_{m,n}|n\rangle_{qq}\langle m|$.

Similar to the steady case, it is necessary to express the above solution in
the real-photon number representation. To this end, one needs the second
time-dependent gauge transformation
\begin{equation*}
U_{2}(t)=e^{\beta _{+}(t)S_{+}}e^{\beta _{-}(t)S_{-}}
\end{equation*}
where the initial condition is taken as $U_{2}(0)=1$. Under the
diagonalization conditions
\begin{eqnarray}
\frac{d\beta _{+}(t)}{dt} &=&\frac{-\varsigma }{4}(1-\beta _{+}(t)^{2})
\notag \\
\frac{d\beta _{-}(t)}{dt} &=&\frac{-\varsigma }{4}(1+2\beta _{+}(t)\beta
_{-}(t)),  \label{beta}
\end{eqnarray}
the rate operator $\tilde{\Gamma}^{\prime }(t)$ (thus $\Gamma (t)$) can be
diagonalized in the real photon representation and reads
\begin{equation*}
\tilde{\Gamma}=U_{2}^{-1}\left( t\right) \tilde{\Gamma}^{\prime
}U_{2}(t)-U_{2}^{-1}\left( t\right) \frac{dU_{2}(t)}{dt}=\frac{-\zeta }{2}
[N-\beta _{+}(t)S_{0}]+\frac{\varsigma }{2}.
\end{equation*}
After the two inverse transformations of $U_{1}(t)$ and $U_{2}(t)$, the
time-dependent solution of Eq. (\ref{masa}) is then
\begin{eqnarray}
\rho (t) &=&U_{1}(t)U_{2}(t)\tilde{\rho}(t)  \notag \\
&=&U_{1}(t)U_{2}(t)\sum_{\substack{ m_{1},n_{1}  \\ n_{2},m_{2}}}
C_{n_{1},n_{2},m_{1},m_{2}}e^{\int_{0}^{t}\{\frac{-\varsigma }{4}[(1-\beta
_{+}(\tau ))(n_{1}+m_{1}+1)+(1+\beta _{+}(\tau ))(n_{2}+m_{2}+1)]+\frac{%
\varsigma }{2}\}d\tau }|n_{1}n_{2}\rangle \langle m_{1}m_{2}|,  \label{solut}
\end{eqnarray}
where we have set the initial state of the system as $\rho (0)=\tilde{ \rho}
(0)=\sum_{m_{1},n_{1},n_{2},m_{2}}C_{n_{1},n_{2},m_{1},m_{2}}|n_{1}n_{2}%
\rangle \langle m_{1}m_{2}|$ in the real photon number representation. This
solution is our central result in the present work. In the following we
consider an explicit initial state to study explicitly the properties of
this solution.

\section{Application of the solution}

In this section, we first consider explicitly the applications of the
solution obtained above to the single- and two-photon processes, which show
some interesting properties. Then we give a brief discussion on the
implication to the quantum communication and quantum computation.

\subsection{Single-photon process}

Consider an initial state as $\rho (0)=|\chi \rangle \langle \chi |$, where $%
|\chi \rangle =a|01\rangle +\sqrt{1-a^{2}}|10\rangle $, and $a$ denotes the
relevant amplitude. In terms of Eqs. (\ref{ac1}, \ref{ac2}), the
time-dependent solution (\ref{solut}) can be written as
\begin{equation}
\rho (t)=\rho _{01,01}(t)|01\rangle \langle 01|+\rho _{10,10}(t)|10\rangle
\langle 10|+\rho _{off}(t)[|10\rangle \langle 01|+|01\rangle \langle
10|]+\rho _{00,00}(t)|00\rangle \langle 00|,
\end{equation}%
where the time-dependent coefficients can be obtained analytically,
\begin{eqnarray}
\rho _{01,01}(t) &=&a^{2}e^{^{\int_{0}^{t}-\frac{\zeta (\beta _{+}(\tau )+1)%
}{2}d\tau }}+(1-a^{2})\beta _{-}(t)^{2}e^{^{\int_{0}^{t}\frac{\zeta (\beta
_{+}(\tau )-1)}{2}d\tau }}+2a\sqrt{1-a^{2}}\beta _{-}(t)e^{^{-\frac{\zeta t}{%
2}}},  \notag \\
\rho _{10,10}(t) &=&a^{2}\beta _{+}(t)^{2}e^{^{\int_{0}^{t}-\frac{\zeta
(\beta _{+}(\tau )+1)}{2}d\tau }}+(1-a^{2})[1+\beta _{+}(t)\beta
_{-}(t)]^{2}e^{^{\int_{0}^{t}\frac{\zeta (\beta _{+}(\tau )-1)}{2}d\tau }}+2a%
\sqrt{1-a^{2}}\beta _{+}(t)[1+\beta _{+}(t)\beta _{-}(t)]e^{^{-\frac{\zeta t%
}{2}}},  \notag \\
\rho _{off}(t) &=&a^{2}\beta _{+}(t)e^{^{\int_{0}^{t}-\frac{\zeta (\beta
_{+}(\tau )+1)}{2}d\tau }}+(1-a^{2})[1+\beta _{+}(t)\beta _{-}(t)]\beta
_{-}(t)e^{^{\int_{0}^{t}\frac{\zeta (\beta _{+}(\tau )-1)}{2}d\tau }}+a\sqrt{%
1-a^{2}}[1+2\beta _{+}(t)\beta _{-}(t)]e^{^{-\frac{\zeta t}{2}}},  \notag \\
\rho _{00,00}(t) &=&\frac{\alpha _{-}(t)}{2}\{a^{2}[1+\beta
_{+}(t)]^{2}e^{^{\int_{0}^{t}-\frac{\zeta (\beta _{+}(\tau )+1)}{2}d\tau
}}+(1-a^{2})[1+\beta _{-}(t)+\beta _{+}(t)\beta _{-}(t)]^{2}e^{^{\int_{0}^{t}%
\frac{\zeta (\beta _{+}(\tau )-1)}{2}d\tau }}  \notag \\
&&+2a\sqrt{1-a^{2}}[1+\beta _{+}(t)][1+\beta _{-}(t)+\beta _{+}(t)\beta
_{-}(t)]e^{^{-\frac{\zeta t}{2}}}\}.  \label{final}
\end{eqnarray}%
From the above solution we can see that $U_{2}(t)$ makes the re-distribution
of the photon number between the two cavities, which produces entanglement
between them, while $U_{1}(t)$ induces de-excitation of the collective modes
and leads to the zero collective mode. In order to study explicitly the time
evolution of the system, it is needed to solve the time-dependent
transformation parameters $\alpha _{-}(t)$ and $\beta _{\pm }(t)$, which can
be obtained from Eq. (\ref{condit2}) and Eqs. (\ref{beta}),
\begin{eqnarray}
\alpha _{-}(t) &=&e^{\varsigma t}-1,  \notag \\
\beta _{+}(t) &=&-\tanh (\frac{\varsigma t}{4}),  \notag \\
\beta _{-}(t) &=&-\frac{1}{2}\sinh (\frac{\varsigma t}{2}).
\label{paremeters}
\end{eqnarray}%
It is known that the entanglement between the two subsystems can be measured
by a quantity named as the entanglement of formation \cite{Bennett}. This
quantity can also be calculated by concurrence $C(\rho )$ \cite{Wootters}.
As usual, $C(\rho )$ is defined as
\begin{equation*}
C(\rho )=\max (0,\lambda _{1}-\lambda _{2}-\lambda _{3}-\lambda _{4}),
\end{equation*}%
where $\lambda _{i}(i=1,\cdots ,4)$ are eigenvalues of matrix
\begin{equation*}
\lbrack \rho ^{1/2}(\sigma _{y}\otimes \sigma _{y})\rho ^{\ast }(\sigma
_{y}\otimes \sigma _{y})\rho ^{1/2}]^{1/2}
\end{equation*}%
and $\lambda _{1}>\lambda _{2}>\lambda _{3}>\lambda _{4}$. Here $\sigma
_{y}=\left(
\begin{array}{cc}
0 & -i \\
i & 0%
\end{array}%
\right) $ and $\rho ^{\ast }$ is the complex conjugation of $\rho $ relative
to the eigenbasis of $\sigma _{z}$. It can be shown that the concurrence
varies from $C=0$ for a disentangled state to $C=1$ for a maximally
entangled state. According to the solution (\ref{final}) it is easy to find
that the concurrence $C(\rho )=2|\rho _{off}(t)|$ in the present case.

In the following we will consider three cases of the entanglement and
decoherence dynamics. For the first case we show that, for product state,
the steady entanglement between the two subsystems can be produced by the
environment. In the second case we present the decoherence-free state which
is stable against the dissipation. The last case represents the situation
where the environment plays the conventional role on system, i.e. it induces
decoherence and makes the system disentangled.

\textit{Case 1, $a=0$}.

In this case the initial state is a product state and has no entanglement.
Substituting Eqs. (\ref{paremeters}) into Eq. (\ref{final}) we have
\begin{equation*}
\rho (t)=\frac{(1-e^{-\frac{\varsigma t}{2}})^{2}}{4}|01\rangle \langle 01|+%
\frac{(1+e^{-\frac{\varsigma t}{2}})^{2}}{4}|10\rangle \langle 10|+\frac{%
e^{-\varsigma t}-1}{4}[|01\rangle \langle 10|+|10\rangle \langle 01|]+\frac{%
1-e^{-\varsigma t}}{2}|00\rangle \langle 00|.
\end{equation*}%
After a long time-evolution, it is noted that all time-dependent
coefficients become constant, and the density matrix can be denoted as
\begin{eqnarray}
\rho _{s} &=&\left(
\begin{array}{cccc}
\frac{1}{2} & 0 & 0 & 0 \\
0 & \frac{1}{4} & -\frac{1}{4} & 0 \\
0 & -\frac{1}{4} & \frac{1}{4} & 0 \\
0 & 0 & 0 & 0%
\end{array}%
\right) =\frac{1}{2}\rho _{0}^{(0)}+\frac{1}{2}\rho _{0}^{(1)},  \notag \\
\rho _{0}^{(0)} &=&|00\rangle \langle 00|,\text{ \ \ }\rho _{0}^{(1)}=|\phi
\rangle \langle \phi |  \label{mar}
\end{eqnarray}%
which is in the basis of $\{|00\rangle ,|01\rangle ,|10\rangle
,|11\rangle \} $ and $\rho _{0}^{(i)}(i=0,1)$ are obtained in
section (IV. A) as zero-mode eigen solutions for zero- and single-
photon processes,
respectively. One can evaluate the concurrence of this steady state as $%
C(\rho )=0.5$. The steady state is the possible mixture of the degenerate
zero-mode eigen solutions. This is consistent with the discussion in section
(IV. A). From the purification scheme proposed in Ref. \cite{Bennett} one
can get the maximal entangled state $|\phi \rangle $ from a collective pairs
of the state above with probability $1/16$. Thus it is shown that the
maximal entangled state $|\phi \rangle $ between the two cavities can be
produced from a product state by the environment.

The time evolution behaviors of the density matrix and the concurrence are
shown in Fig. 1 and 2, respectively.

The same discussion can be applied to the initial product states with $a=\pm
1$, in which $\rho (t)$ and $C(\rho)$ have the similar time evolution
behaviors.

\textit{Case 2, $a=-1/\sqrt{2}$}

The initial state is the maximally entangled state. In this case we found
that the density matrix is time-independent, i.e.,
\begin{equation*}
\rho (t)=\frac{1}{4}|01\rangle \langle 01|+\frac{1}{4}|10\rangle \langle 10|-%
\frac{1}{4}[|01\rangle \langle 10|+|10\rangle \langle 01|]=|\phi \rangle
\langle \phi |.
\end{equation*}
Thus, this solution is decoherence-free and the concurrence is
also a maximal constant $C(\rho )=1$. From section (IV. A) we know
that $|\phi \rangle \langle \phi |$ is the zero-mode eigen
solution for single-photon process of the master equation
(\ref{masa}).

\textit{Case 3, $a=1/\sqrt{2}$}

In this case we obtain the time dependent solution as follows
\begin{equation*}
\rho(t)=\frac{e^{-\varsigma t}}{2}\{|01\rangle \langle 01|+|10\rangle
\langle 10|+[|01\rangle \langle 10|+|10\rangle \langle
01|]\}+(1-e^{-\varsigma t})|00\rangle \langle 00|.
\end{equation*}
Its time evolution is plotted in Fig. 3, and asymptotically the state
approach to a product state with zero real photon number.
\begin{equation*}
\rho _{s}=\rho _{0}^{(0)}.
\end{equation*}
The concurrence $C(\rho )$ decays quickly, as shown in Fig. 4, indicating
that the decoherence is very strong and the two cavities will lose all of
the information contained in its entangled initial state.

Except for the above three special cases, it is difficult to obtain a simple
expression of the density matrix for general cases of $a$ and numerical
calculations are needed. In Fig. 5, we show the time evolution of the
density matrix for a general initial state characterized by $a$. Fig. 6
presents the time behavior of the corresponding concurrence.

\subsection{Two-photon process}

In this two-photon process, each mode of the two cavities is
related to three states, $|0\rangle $, $|1\rangle $, and
$|2\rangle$, i.e., a qutrit. Therefore, this system is a potential
candidate for a qutrit quantum information processing
\cite{Alber,Bruss}. The advantage of qutrits instead of qubits is
that the qutrits are much secure against the symmetric attacks in
a quantum key distribution protocol, as shown by Bruss and
Macchiavello \cite{Bruss}. In a recent experiment \cite{Bogdanov},
the arbitrary qutrit states have been realized on the single-mode
biphoton field, which shares some similarities with our model
considered here. Based on the exact solution Eq. (\ref{solut}) in
the two-photon process we shall analyse the decoherence effects on
the qutrit state generation.

Following Ref. \cite{Cereceda}, the entanglement of bipartite qutrit states
can be measured in terms of the Schmidt coefficients by
\begin{eqnarray}
|\Phi _{3\times 3}\rangle &=&k_{1}|x_{1},y_{1}\rangle
+k_{2}|x_{2},y_{2}\rangle +k_{3}|x_{3},y_{3}\rangle ,  \notag \\
C(|\Phi _{3\times 3}\rangle ) &=&\sqrt{
3(k_{1}^{2}k_{2}^{2}+k_{1}^{2}k_{3}^{2}+k_{2}^{2}k_{3}^{2})},  \label{3con}
\end{eqnarray}
where Eq. (\ref{3con}) is a generalized definition of the concurrence for
qutrit states.

Suppose that the initial pure state is $\rho (0)=|\omega \rangle \langle
\omega |$ , where $|\omega \rangle =a|02\rangle +b|11\rangle +c|20\rangle $
where the coefficients $a, b$, and $c$ satisfy $\sqrt{ a^{2}+b^{2}+c^{2}}=1$%
. It is very difficult to discuss an arbitrary case. Instead, we focus on
the following special cases.

\textit{Case 1, $a=1,b=c=0$}

The time-dependent solution can be obtained
\begin{eqnarray}
\rho (t) &=&\frac{(e^{-\varsigma t}+1)^{2}}{4}|\varphi (t)\rangle \langle
\varphi (t)|+\frac{1-e^{-2\varsigma t}}{2}|\phi (t)\rangle \langle \phi (t)|+%
\frac{(1-e^{-\varsigma t})^{2}}{4}|00\rangle \langle 00|,  \notag \\
|\varphi (t)\rangle &=&\frac{(1+e^{^{-\varsigma t}/2})^{2}}{2(e^{-\varsigma
t}+1)}|02\rangle -\frac{1-e^{-\varsigma t}}{\sqrt{2}(e^{-\varsigma t}+1)}%
|11\rangle +\frac{(1-e^{^{-\varsigma t}/2})^{2}}{2(e^{-\varsigma t}+1)}%
|20\rangle ,  \notag \\
|\phi (t)\rangle &=&\frac{1+e^{-\varsigma t/2}}{\sqrt{2(e^{-\varsigma t}+1)}}%
|01\rangle -\frac{1-e^{-\varsigma t/2}}{\sqrt{2(e^{-\varsigma t}+1)}}%
|10\rangle .  \label{enst}
\end{eqnarray}%
From Eqs. (\ref{enst}) we can see that with the time evolution $\rho (t)$
tends to $\frac{1}{4}|\varphi \rangle \langle \varphi |+\frac{1}{2}|\phi
\rangle \langle \phi |+\frac{1}{4}|00\rangle \langle 00|$, which is a
mixture of two-photon, one-photon and zero-photon zero-mode solutions, as
analyzed in section (IV. A). Since the quantification of entanglement of
mixed entangled qutrit states is not yet well understood, we can give an
upper bound of the entanglement by the converxity property of entropy
function, namely,
\begin{equation*}
E(\rho (t))\leq \frac{(e^{-\varsigma t}+1)^{2}}{4}E(|\varphi
(t)\rangle \langle \varphi (t)|)+\frac{1-e^{-2\varsigma
t}}{2}E(|\phi (t)\rangle \langle \phi (t)|)+\frac{(1-e^{-\varsigma
t})^{2}}{4}E(|00\rangle \langle 00|)=E^{\ast }(\rho (t)).
\end{equation*}%
In the above expression, $E$ is defined by $E=h(\frac{1+\sqrt{1-C^{2}}}{2})$%
, where $h(x)=-x\log _{2}x-(1-x)\log _{2}(1-x)$ and $C$ is the concurrence
of the related density matrix. From Eq. (\ref{3con}) the two-photon pure
state entanglement $E(|\varphi (t)\rangle \langle \varphi (t)|)$ can be
obtained. $E(|\phi (t)\rangle \langle \phi (t)|)$ is just a pure qubit
entanglement, which can be calculated in the traditional way. Fig. 7 shows
the time evolution of $E^{\ast }(\rho (t))$ (solid line). One notes that the
steady entanglement can be generated from the product initial state.

The similar discussions can also be applied to the cases of $c=1,a=b=0$ and $%
b=1,a=c=0$. While the evolution of the upper bound $E^{\ast }(\rho (t))$ for
$b=1,a=c=0$ is the same as the above, a slightly different evolution in the
case of $c = 1, a = b = 0$ is found, as shown in Fig. 7 by the dashed line.
Nevertheless, the conclusion of the generation of the entanglement is same.

\textit{Case 2, $a=b=\frac{1}{2},c=-1/\sqrt{2}$}

In this case the solution is time-independent,
\begin{equation*}
\rho (t)=|\varphi \rangle \langle \varphi |,
\end{equation*}
where $|\varphi \rangle $ is the same as the form in section (IV. A). Once
again we verify the existence of decoherence-free state.

\textit{Case 3, $a=b=\frac{1}{2},c=1/\sqrt{2}$}

From the analysis of last subsection we know that environment can produce
and destroy entanglement, which is dependent on the initial state. This is
still true in the two-photon process. The \textit{Case 1 }corresponds to the
entanglement production. In this case we shall see the environment can
destroy entanglement in certain initial state. The solution read
\begin{eqnarray*}
\rho (t) &=&e^{-2\varsigma t}|\chi \rangle \langle \chi |+2(e^{-\varsigma
t}-e^{-2\varsigma t})|\upsilon \rangle \langle \upsilon |+(1-e^{-\varsigma
t})^{2}|00\rangle \langle 00|, \\
|\chi \rangle &=&\frac{1}{2}|02\rangle +\frac{1}{\sqrt{2}}|11\rangle +\frac{1%
}{2}|20\rangle , \\
|\upsilon \rangle &=&\frac{1}{\sqrt{2}}|01\rangle +\frac{1}{\sqrt{2}}%
|10\rangle .
\end{eqnarray*}%
We can see $\rho (t)$ tends to zero-photon zero-mode solution $|00\rangle
\langle 00|$. As a result, the entanglement is destroyed completely. The
time-evolution for $E^{\ast }(\rho (t))$ is plotted in Fig. 7, as shown as
dotted line.

The discussion above is limited to very special initial states, which are
formulated in the Fock states representation. It is very interesting to
expand the present discussion to the Gaussian states, which can be created
relatively easily and can be widely applied to quantum cryptography and
quantum teleportation. In addition, the entanglement criteria \cite%
{giedke01, duan00, simon00} and the formation of entanglement \cite{giedke03}
are also available. However, instead of the Fock state representation, it is
very convenient to use the continuous variable representation to discuss the
Gaussian states, which is beyond the present work.

\section{\protect\bigskip Summary}

In summary, we have investigated the master equation of two identical
single-mode cavity fields coupled to a common quantum environment. Using the
algebraic dynamical method, the two kinds of dynamical symmetries, namely
the $sp(4)\supset $ $su(1,1)\supset \{K_{0}\}$ and $sp(4)\supset u(2)$ $%
\supset \{U,S_{0}\}$ dynamical symmetries of the master equation have been
explored. The first dynamical symmetry corresponds to the collective
quasi-photon mode, while the second one is the dynamical symmetry for the
real photons of the two cavities. According to algebraic dynamics, the
system is integrable and its analytical solutions have been obtained. It is
found that the system is affected by the environment through the rate
operator consisting merely of collective quasi-photon operators and having
collective quasi-photon eigen modes which are highly degenerate in terms of
the real photon states. Based on the analytical solutions, the decoherence
and entanglement properties of the quasi-photon zero mode of the rate
operator have been studied in detail, and the surprising finding is that
different entangled states can be produced from the disentangled initial
states by environment effects in some cases. The results obtained are useful
for the encoding scheme in quantum computation and for the entanglement
generation of the 2-qubit and 2-qutrit cavity systems.

\section{\protect\bigskip Acknowledgements}

This work was supported in part by the National Natural Science Foundation
under grants No.10175029,10375039, and 10474103, the Doctoral Education Fund
of the Education Ministry and by the Nuclear Theory Research Fund of HIRFL
of China.

\newpage

\begin{figure}[tbp]
\scalebox{1.0}{\includegraphics{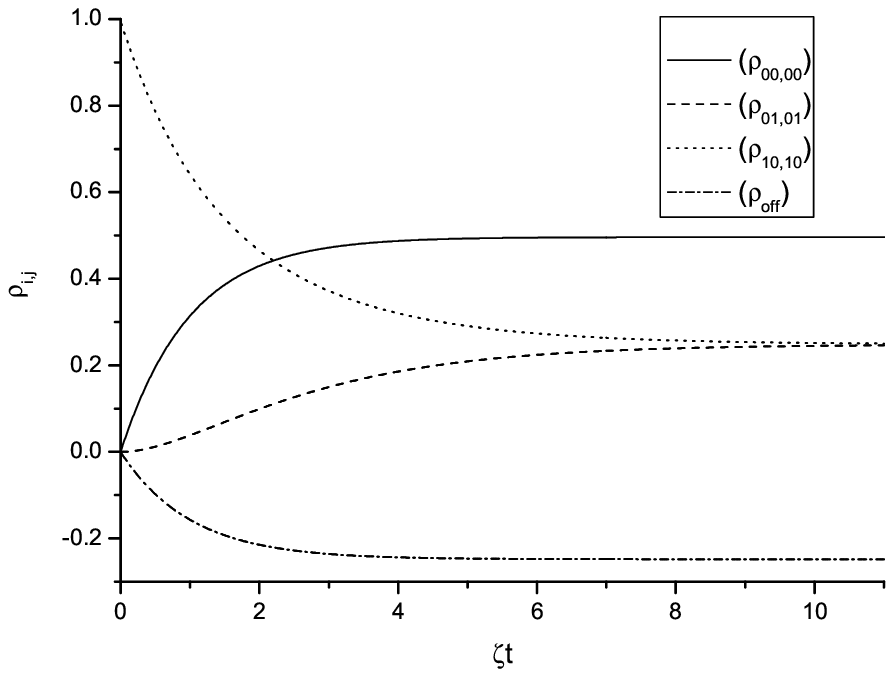}} \caption{The time
evolution of the elements of density matrix with the initial
condition of $a=0$.} \label{Fig1}
\end{figure}

\begin{figure}[tbp]
\scalebox{1.0}{\includegraphics{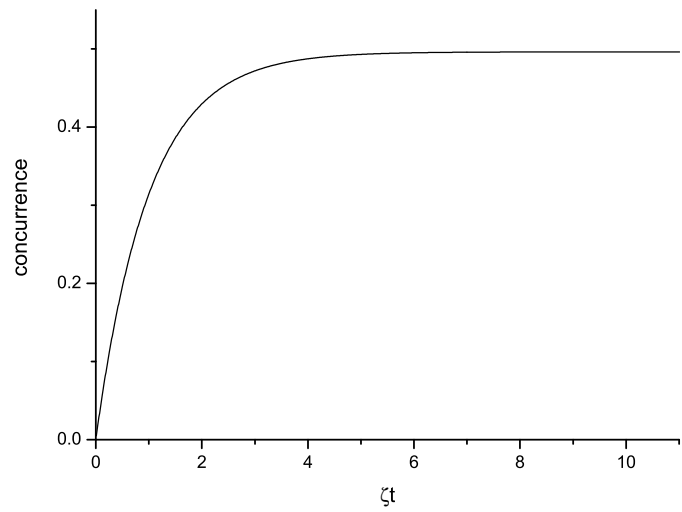}} \caption{The time
evolution of the concurrence $C(\protect\rho )$ with the initial
condition of $a=0$.} \label{Fig2}
\end{figure}

\begin{figure}[tbp]
\scalebox{1.0}{\includegraphics{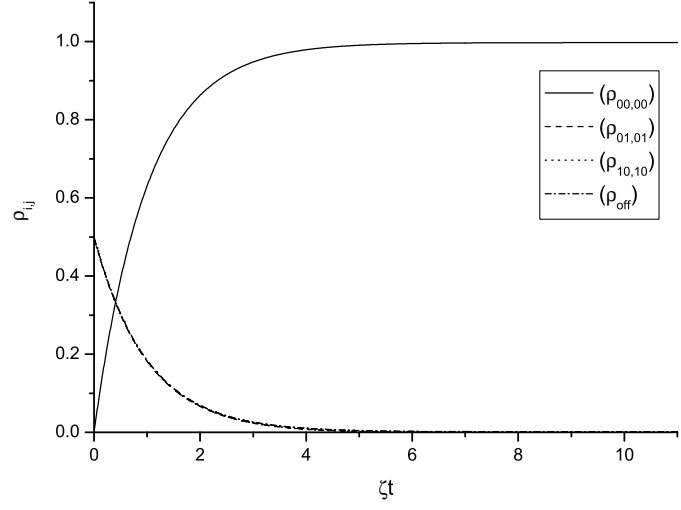}} \caption{The time
evolution of the elements of density matrix with the initial
condition of $a=1/\protect\sqrt{2}$.} \label{Fig3}
\end{figure}

\begin{figure}[tbp]
\scalebox{1.0}{\includegraphics{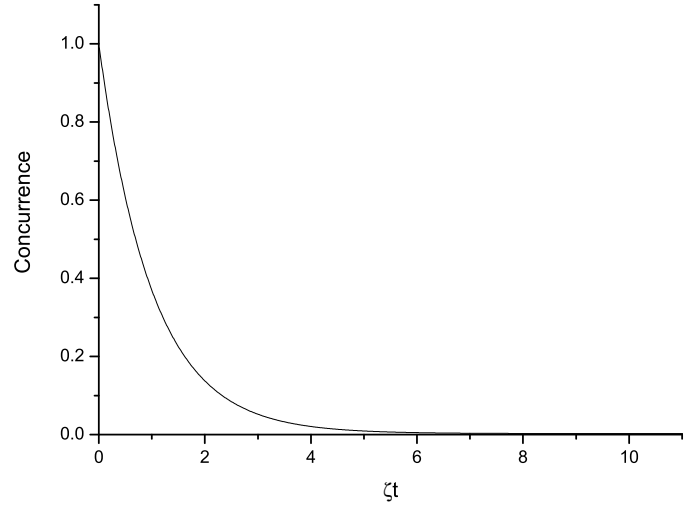}} \caption{The time
evolution of the concurrence $C(\protect\rho )$ with the initial
condition of $a=1/\protect\sqrt{2}$.} \label{Fig4}
\end{figure}

\begin{figure}[tbp]
\scalebox{0.6}{\includegraphics{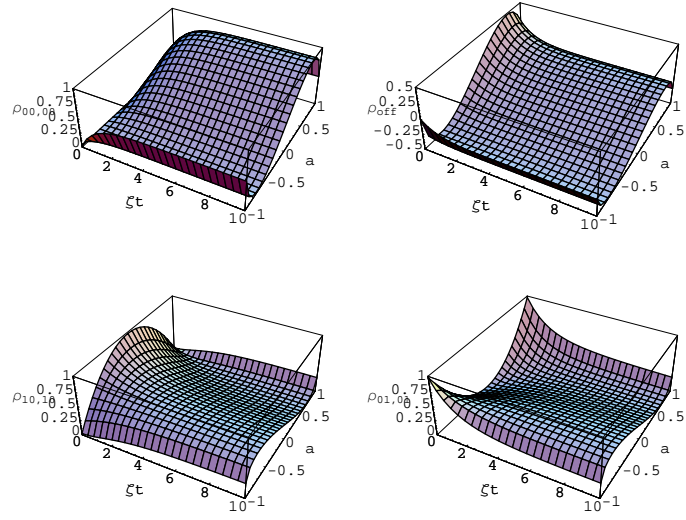}} \caption{The time
evolution of the elements of density matrix with a general initial
condition given by $a$.} \label{Fig5}
\end{figure}

\begin{figure}[tbp]
\scalebox{0.8}{\includegraphics{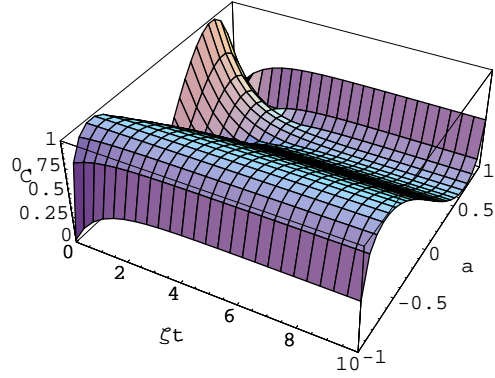}} \caption{The time
evolution of the concurrence $C(\protect\rho )$ with different $a$.}
\label{Fig6}
\end{figure}

\begin{figure}[tbp]
\scalebox{0.8}{\includegraphics{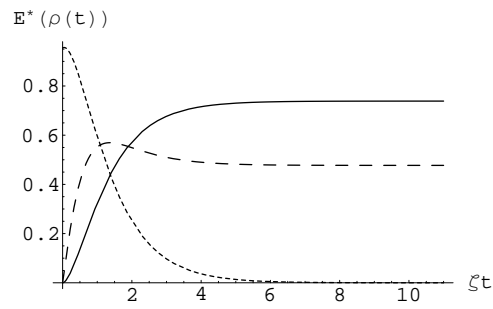}} \caption{The upper bound
$E^{\ast }$ of the entanglement of two-photon process for different
cases: $a = 1, b = c = 0$ (solid line), $c = 1, a = b = 0$ (dashed
line), and $a = b = \frac{1}{2}, c = 1/\protect\sqrt{2}$ (dotted
line). } \label{Fig7}
\end{figure}

\end{document}